\documentclass[10pt, doublecolumn]{IEEEtran}
\usepackage{epsfig,latexsym}
\usepackage{float}
\usepackage{indentfirst}
\usepackage{amsmath}
\usepackage{amssymb}
\usepackage{times}
\usepackage{subfigure}
\usepackage{psfrag}
\usepackage{hyperref}
\usepackage{cite}
\usepackage{lastpage}
\usepackage{fancyhdr}
\usepackage{color}
 \usepackage{amsthm}
\usepackage{bigints}
\newcounter{problem}
\newcounter{save@equation}
\newcounter{save@problem}
\makeatletter

\begin{document}
\title{ \vspace{-0.5em}{\huge Unveiling  the Importance  of SIC in NOMA Systems:
\\ Part I - State of the Art and Recent Findings }\\\vspace{-0.25em}
{\Large {\it (Invited Paper)}}}\vspace{-0.25em}

\author{ Zhiguo Ding, \IEEEmembership{Fellow, IEEE}, Robert Schober, \IEEEmembership{Fellow, IEEE}, and H. Vincent Poor, \IEEEmembership{Life Fellow, IEEE}    \thanks{ 
  
\vspace{-2em}

    Z. Ding and H. V. Poor are  with the Department of
Electrical Engineering, Princeton University, Princeton, NJ 08544,
USA. Z. Ding
 is also  with the School of
Electrical and Electronic Engineering, the University of Manchester, Manchester, UK (email: \href{mailto:zhiguo.ding@manchester.ac.uk}{zhiguo.ding@manchester.ac.uk}, \href{mailto:poor@princeton.edu}{poor@princeton.edu}).
R. Schober is with the Institute for Digital Communications,
Friedrich-Alexander-University Erlangen-Nurnberg (FAU), Germany (email: \href{mailto:robert.schober@fau.de}{robert.schober@fau.de}).

  }\vspace{-2em}}
 \maketitle
 
\begin{abstract}  
The key idea of non-orthogonal multiple access (NOMA) is to serve multiple users simultaneously at the same time and frequency, which can  result in excessive multiple-access interference.  As a crucial component of NOMA systems, successive interference cancelation (SIC) is key to combating this multiple-access interference, and is focused on in this letter,  where an overview of       SIC decoding order selection schemes   is provided. In particular, selecting  the SIC decoding order based on  the users' channel state information (CSI)    and   the users' quality of service (QoS), respectively, is discussed.   The limitations of these  two approaches   are illustrated, and then  a recently proposed scheme, termed hybrid SIC, which   dynamically adapts the SIC decoding order is presented and shown to achieve a surprising performance improvement that cannot be realized by   the conventional SIC decoding order selection  schemes  individually.
\end{abstract} \vspace{-1.2em} 

\section{Introduction}
As a paradigm shift for the design of multiple access techniques, non-orthogonal multiple access (NOMA)  encourages spectrum sharing among multiple users, instead of forcing  them to individually  occupy orthogonal resource    blocks as in conventional orthogonal multiple access (OMA) \cite{mojobabook}. As a result, NOMA can significantly improve   spectral efficiency, reduce access delay,  and support massive connectivity.  As one of the most promising   multiple access  techniques,  NOMA has   been extensively  studied under the   3rd Generation Partnership Project (3GPP)  framework, from Release 14 in 2015 to Release 16 in 2019, where NOMA was formally adopted for downlink transmission  in Release 15, also termed Evolved Universal Terrestrial Radio Access (E-UTRA) \cite{3gpp1,3gpp15,3gpp3}.  With the current  rollout of the fifth-generation (5G)   wireless systems, significant efforts are being made towards  the full inclusion of NOMA in beyond 5G   systems  \cite{8972353}. 

A unique feature of NOMA systems is the existence of excessive multiple-access interference, which is  caused by the spectrum sharing among the users. Successive interference cancelation (SIC) has been shown to be an effective method  to combat this interference   \cite{6692307,Verduebook}. Due to the sequential nature of SIC, the SIC decoding order is a key issue in the implementation of SIC, and will be focused on in this two-part invited paper.   The first part of the   paper aims to provide an overview for how the SIC decoding order has been determined in NOMA, and to illustrate its impact on the performance of NOMA. In particular, selecting the SIC decoding order  based on   the users' channel state information (CSI)     is considered  first, since this is  a straightforward choice   and has been used since the invention of NOMA \cite{6692307,Nomading}. Then, selecting the SIC decoding order based on     the users' quality of service (QoS) requirements   is  considered,   the rationale behind this approach   is explained,  and    ideal application scenarios for it  are illustrated \cite{Zhiguo_CRconoma,8352630}. We note that in most existing NOMA works, the SIC decoding order is prefixed and based on either of the two aforementioned criteria, which may suggest   that  swapping SIC decoding orders is trivial and does not yield   a significant performance gain. However, a recent work      \cite{SGFx} has shown the opposite to be true. In fact,  dynamically switching the SIC decoding order  can achieve a surprising performance improvement that cannot be realized by the two conventional   schemes. This improvement is illustrated  in this paper, and the underlying reasons   are   explained in detail.   
\vspace{-0.5em}
 \section{NOMA Using  CSI-Based SIC Decoding Order }
For purposes of illustration, this letter  considers an uplink communication scenario with $(M+1)$ users, denoted by   ${\rm U}_m$, $0\leq m\leq M$,  where ${\rm U}_m$'s channel     gain is denoted by $h_m$.  Using the users' channel conditions to select  the SIC decoding order is a straightforward choice and has been adopted in many forms of NOMA \cite{6692307,Nomading}. Take two-user power-domain NOMA  as an example, which can serve two users with different channel conditions  simultaneously. Without loss of generality, assume that ${\rm U}_n$, $1\leq n \leq M$, is scheduled and paired with ${\rm U}_0$ for NOMA transmission, under the condition that   $ |h_n|^2\geq 
|h_0|^2$.
 
Conventional  OMA  serves the two users in different resource blocks, and the users' data rates  are $R_i^O=\frac{1}{2}\log\left(1+P^O|h_i|^2\right)$, $i\in\{0,n\}$, respectively, where  the users' transmit powers are assumed to be identical  and denoted by $P^O$. The shortcomings  of OMA can be illustrated by considering the extreme case $h_0\rightarrow 0$, i.e., ${\rm U}_0$ experiences deep fading. As a result, the resource block allocated to ${\rm U}_0$   is wasted due to the user's poor channel conditions.   

Power-domain NOMA  serves  the two users   simultaneously, and applies SIC at the base station for signal separation.     A natural SIC strategy is to first decode the signal from  the strong user, ${\rm U}_n$, and then decode ${\rm U}_0$'s  signal if ${\rm U}_n$'s signal can be decoded and removed successfully, which yields   the following achievable data rates:
\begin{align}
R_n^N =  \log\left(1+\frac{P|h_n|^2}{1+P|h_0|^2}\right),
\end{align}
and 
\begin{align}
R_0^N = \log\left(1+P|h_0|^2\right),
\end{align}
respectively, where   the users'   transmit powers   are also assumed to be identical  and equal to  $P$. The benefits of NOMA   can be clearly demonstrated  by again considering  the extreme case $h_0\rightarrow 0$. Assume  $P=\frac{1}{2}P^O$ for a fair comparison between OMA and NOMA. The   sum rate of NOMA can be approximated as follows:
\begin{align}\label{appro1}
R_{sum}^N =R_0^N+R_n^N  \underset{h_0\rightarrow 0}{\longrightarrow} \log\left(1+ P|h_n|^2 \right)\underset{P\rightarrow \infty}{\longrightarrow} \log P,
\end{align}
which is almost   two times   the sum rate of OMA, $R_{sum}^O \triangleq R_0^O+R_n^O   \rightarrow \frac{1}{2}\log P$,   where we assume that $P|h_0|^2\rightarrow 0$. 

{\it Remark 1:} As the number of users   participating in power-domain NOMA grows, all   users except the one whose signal is decoded at the last stage of SIC   suffer from severe  interference, which means that  their QoS experience deteriorates. Thus,  it is difficult to apply    power-domain NOMA to a general case with more than two users,     which is a disadvantage of power-domain NOMA compared to other forms of NOMA.   

{\it Remark 2:}  Another disadvantage of   power-domain NOMA is that the users' channel conditions need to be sufficiently different in order to yield a reasonable performance gain over OMA, which can be illustrated by considering  the special case   $h_0=h_n$ and $P=\frac{1}{2}P^O$. It is straightforward to show that, in this case, $R_{sum}^N=R_{sum}^O =  \log\left(1+2P|h_0|^2\right)$. In other words, if the  users' channel qualities  are identical, the performance of NOMA is the same as that of OMA, but the system complexity of NOMA is higher than that of OMA.     These   disadvantages can be avoided by using QoS-based SIC. 

 \section{NOMA Using  QoS-Based  SIC Decoding Order }\label{section qos}
Cognitive-radio inspired NOMA (CR-NOMA) is a well-known example  for using the users' QoS requirements to select the SIC order \cite{Zhiguo_CRconoma,8352630}. 
Unlike power-domain NOMA,  in CR-NOMA,      ${\rm U}_0$ is assumed to be a delay-sensitive user with a low target data rate, denoted by ${R}_0$, e.g.,   ${\rm U}_0$ may be a voice-call user or   an Internet of Things (IoT) healthcare device which needs to send urgent health status changes. On the other hand, it is assumed that ${\rm U}_n$, $1\leq n \leq M$, can be served in a delay tolerant manner, e.g., ${\rm U}_n$ may be  a peer-to-peer file sharing  user  or an IoT device   sending   personal health records.    In   OMA, because of its delay-sensitivity, ${\rm U}_0$ is allowed to occupy a dedicated   resource block, such as a time slot, which results in low spectral efficiency since ${\rm U}_0$ has only a small amount of data to be delivered to the base station. 

The key idea of CR-NOMA is to treat NOMA as a special case of a CR system, where ${\rm U}_0$ is viewed as the primary user and ${\rm U}_n$, $1\leq n \leq M$, are viewed as secondary users.  CR-NOMA   ensures that   the secondary users are admitted   to   the channel, which in OMA would be solely occupied by ${\rm U}_0$, while guaranteeing    ${\rm U}_0$'s QoS requirements. 
In CR-NOMA, the primary user's signal is decoded first. For illustration purposes, we assume  that a single secondary user,   ${\rm U}_n$, $1\leq n\leq M$, is scheduled based on the following metric: 
\begin{align}\label{cons2}
 \log\left(1+\frac{ P|h_0|^2}{1+P|h_n|^2}\right)\geq  {R}_0,
\end{align}
in order to guarantee   ${\rm U}_0$'s QoS requirements. 
If the constraint in \eqref{cons2} is feasible, the first stage of SIC is guaranteed to be successful, and  ${\rm U}_n$'s achievable data rate is given by
\begin{align}\label{cons3}
R_n^{CR}=\log\left(1+P|h_n|^2\right).
\end{align}
If   none of the secondary users can satisfy the constraint in \eqref{cons2},   OMA  is used, in order to avoid any performance degradation for ${\rm U}_0$. In other words, the use of CR-NOMA   is transparent to ${\rm U}_0$. 

\subsubsection{Rationale behind the used SIC order} CR-NOMA  decodes first the signals from ${\rm U}_0$, the user with the low data rate requirement. 
The rationale behind  this SIC order is two-fold. Firstly, the user whose signals are decoded in the first stage of SIC suffers strong interference, as can be observed from \eqref{cons2}, which means that the user's achievable data rate will be small. This is not an issue, since ${\rm U}_0$'s  target data rate is assumed   to be not demanding. Secondly, since ${\rm U}_n$'s signals are decoded in the last stage of SIC, they do not suffer from any interference, as can be observed from \eqref{cons3}. Recall that ${\rm U}_n$'s data rate constitutes  the performance gain of CR-NOMA over OMA. Therefore, the fact that there is no interference in \eqref{cons3} promises a significant performance gain of NOMA over OMA, as discussed in the following subsection.  

\subsubsection{The benefits   of QoS-based SIC} We use   two examples to illustrate the benefits to use QoS-based SIC.  The first example is to show that   QoS-based SIC can be easily extended to a general case with more than two users, while guaranteeing the users' QoS requirements. In particular,  we  assume that  there are $M$   delay-sensitive  users to be served at low data rates, and one delay-tolarent user. In OMA, $(M+1)$ time slots are needed to serve these users.   By using NOMA with QoS-based SIC, it is possible to serve all users  in a single time slot, which means that the spectral efficiency can be improved $(M+1)$ times compared to OMA. This benefit is particularly important for the application of NOMA in the context of massive multiple access, where massive connectivity has to be provided to reduce the access delay in IoT networks \cite{9031550}.  

For the second example, we consider   the special  case, where  ${\rm U}_0$'s channel gain is the same as ${\rm U}_n$'s, i.e., $ h_0=h_n=h$.     As discussed in Remark 2, the use of NOMA with CSI-based SIC  does not offer any performance gain over OMA if $ h_0=h_n$. With QoS-based SIC, the sum rate gain of NOMA over OMA is simply ${\rm U}_n$'s data rate, if ${\rm U}_n$ can be admitted, i.e., $ \log\left(1+\frac{ P|h|^2}{1+P|h|^2}\right)\geq  {R}_0$, otherwise the performance gain is zero. Therefore, the sum rate gain of NOMA over OMA is given by
\begin{align}\label{gainx}
\Delta_{sum} =& \mathbf{1}_0\log\left(1+P|h|^2\right),
\end{align}
where  $\mathbf{1}_0$ is an indicator function, i.e.,  $\mathbf{1}_0=1$ if $ \log\left(1+\frac{ P|h|^2}{1+P|h|^2}\right)\geq  {R}_0  $, otherwise $\mathbf{1}_0=0$.
Assume that $h$ is a Rayleigh fading channel gain, i.e., $h$ is complex Gaussian distributed with zero mean and unit variance. Then, it is straightforward to show that 
\begin{align}
{\rm P}(\mathbf{1}_0=1) = e^{ - \frac{2^{ {R}_0}-1}{P\left(2-2^{ {R}_0}\right)}}\rightarrow 1,
\end{align}
for $P\rightarrow \infty$, where we assume that $ {R}_0<1$ bit/s/Hz. Therefore, $\Delta_{sum} \rightarrow \infty$, for $P\rightarrow \infty$, which means that NOMA with QoS-based SIC   can   offer a significant performance gain over OMA, even if all users have the same channel condition. In contrast, the performance gain of NOMA with CSI-based SIC  diminishes in this case. This property is particularly important for the application of NOMA in indoor communication environments, where the users' channel conditions are expected to be similar.

\subsubsection{The implications of the channel conditions} Because  the SIC decoding order of CR-NOMA  is not decided by the users' channel conditions, it may happen that   ${\rm U}_0$'s channel conditions are weaker  than   ${\rm U}_n$'s, i.e.,  $|h_0|^2<|h_n|^2$. In other words,  during the first step of SIC, the   signal strength might be weaker than the interference strength,  which leads to the common   question whether this situation  results in a decoding failure. We note that whether a signal can be decoded correctly   depends on whether the data rate supported by the channel is larger than the target data rate, i.e., $ \log\left(1+\frac{ P|h|^2}{1+P|h|^2}\right)\geq  {R}_0  $. As long as this condition holds,  the use of error correction coding can ensure that the signal is correctly decoded,  even if the signal   strength is weaker than the interference strength. Error correction coding  injects redundant information, which reduces the information data rate. But if ${\rm U}_0$'s target data rate is small, a significant amount of redundant information can be added. For example, for the case of $ {R}_0=0.1$ bits/s/Hz, a repetition code with a code rate of $\frac{1}{10}$ is affordable, where one bit is repeated $10$ times. With so much redundant information injected,  signals  can be successfully decoded, even in the presence of strong interference. 

\section{NOMA Using Hybrid SIC with Adaptive Decoding Order }
Hybrid SIC   was originally  proposed for NOMA assisted semi-grant-free (SGF) transmission in \cite{SGFx}. In this section, the motivation for using   hybrid SIC with adaptive decoding order  is   provided first, its key idea is then illustrated for  a general NOMA uplink scenario, and finally its performance is demonstrated by using the CSI and QoS based SIC decoding orders  as the benchmarks. 
\vspace{-0.5em}
\subsection{Limitations of   CSI/QoS-Based SIC}
Without loss of generality, we focus on the same uplink scenario as   in Section \ref{section qos}, i.e.,  one of the $M$ secondary users is   admitted to the channel which would be solely occupied by ${\rm U}_0$  in   OMA  \cite{8662677}.  In addition, we   assume that the secondary users'  channel gains are ordered as $
|h_1|^2\leq\cdots\leq    |h_M|^2$. 

\begin{figure}[!t]\centering  
    \epsfig{file=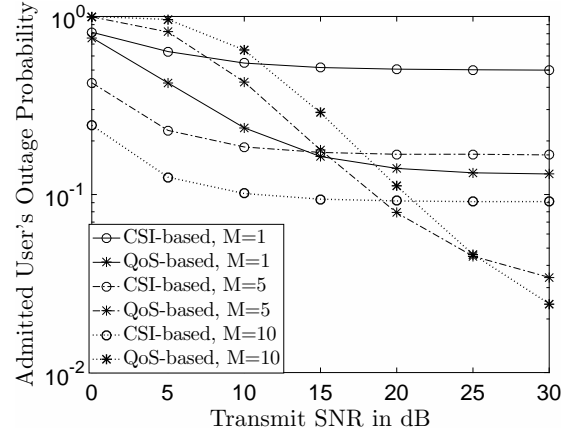, width=0.4\textwidth, clip=}\vspace{-0.5em}
\caption{ Outage performance achieved by NOMA transmission with the two types of SIC. Independent and identically distributed (i.i.d.) Rayleigh fading is assumed for the users' channel gains.   $R_0=0.2$ bits/s/Hz, and $R_s=1$ bits/s/Hz. \vspace{-1em} }\label{fig1}\vspace{-1em}
\end{figure}

NOMA with CSI-based SIC  decodes the strong user's signal first. One possible scheme, termed SGF Scheme I in \cite{8662677}, is to schedule the user with the strongest channel gain among the  $M$ secondary users, i.e., ${\rm U}_M$, and require  the base station to decode ${\rm U}_M$'s signal at the first stage of SIC.  In order to guarantee that ${\rm U}_0$ experiences the same QoS as in OMA,   ${\rm U}_M$ needs to use   the following   data rate for its transmission 
\begin{align}\label{csirate}
R^{CSI}_M =  \log\left(1+\frac{P|h_M|^2}{1+P|h_0|^2}\right),
\end{align}
  which   guarantees that the first stage of SIC can be carried out successfully.  Therefore, at the second stage of SIC, ${\rm U}_0$'s signal can be  decoded without suffering any interference. In other words, an additional user, ${\rm U}_M$, is admitted to the channel, while ${\rm U}_0$ transmits  as if it solely occupied the channel, which is an  advantage of this scheme. In addition, this scheme can efficiently exploit  multi-user diversity, i.e.,    increasing $M$ can reduce  the admitted user's outage probability, defined by ${\rm P}^{CSI}\triangleq {\rm P} \left(  R^{CSI}_M< {R}_s\right)$, as shown in Fig.~\ref{fig1}. 

A disadvantage of this scheme is that   there is an error floor for ${\rm U}_M$'s outage probability. In particular, ${\rm P}^{CSI}$ can be approximated as follows:
\begin{align}\nonumber 
{\rm P}^{CSI}& = {\rm P} \left(   \log\left(1+\frac{P|h_M|^2}{1+P|h_0|^2}\right)< {R}_s\right)
\\\label{appr1}&
\underset{P\rightarrow \infty}{\rightarrow} {\rm P} \left(   \log\left(1+\frac{|h_M|^2}{|h_0|^2}\right)< {R}_s\right),
\end{align}
which is a constant and not a function of the transmit signal-to-noise ratio (SNR), where the approximation is obtained for   $P\rightarrow \infty$ and we assume that all secondary users have the same target data rate, denoted by $R_s$. This error floor can potentially lead to a degradation of transmission robustness. In addition, the approximation in \eqref{appr1} indicates that ${\rm U}_M$'s data rate is   capped   at high SNR. Therefore,   the performance gain of NOMA over OMA  is also capped since this gain   is related to  the admitted user's data rate as shown in \eqref{gainx}.

NOMA with QoS-based SIC    first decodes the signal from   primary user, ${\rm U}_0$, by treating   the admitted secondary user's signal  as noise.   In order to minimize the performance degradation of ${\rm U}_0$, one possible scheme,  termed SGF Scheme II in \cite{8662677}, is to schedule the secondary user with the weakest channel gain, i.e.,    ${\rm U}_1$, which yields the following   data rate:
\begin{align}\label{qosrate}
R^{QoS}_1 = \log(1+|h_1|^2P),
\end{align}
if  $  \log\left(1+\frac{P|h_0|^2}{1+P|h_1|^2}\right)> {R}_0$, otherwise $R^{QoS}_1=0$. Therefore,   ${\rm U}_1$'s outage probability is given by
 \begin{align}
&{\rm P}^{QoS} = {\rm P} \left(   \log\left(1+\frac{P|h_0|^2}{1+P|h_1|^2}\right)< {R}_0\right)
\\\nonumber
 & + {\rm P} \left(   \log\left(1+\frac{P|h_0|^2}{1+P|h_1|^2}\right)> {R}_0, \log\left(1+ P|h_1|^2 < {R}_s\right)\right).
\end{align}
An  advantage of this scheme is that the admitted user's signal is  interference free, as is evident  from \eqref{qosrate}, which means that the performance gain of NOMA over OMA is not capped, unlike in CSI-based SIC. This is also the reason why   QoS-based SIC outperforms   CSI-based SIC at high SNR  in Fig. \ref{fig1}.  A disadvantage of  QoS-based SIC is that it cannot efficiently use multi-user diversity. For example, Fig. \ref{fig1} shows that    increasing $M$ deteriorates   its outage probability, since the channel gain $|h_1|^2$ becomes weaker as $M$ increases.   Another disadvantage of   QoS-based SIC   is that an outage probability error floor still exists, as shown in Fig. \ref{fig1}.  
\vspace{-1em}
\subsection{NOMA with Hybrid SIC }
The discussions in the previous subsection suggest that realizing  NOMA transmission without outage probability error floors is a mission impossible. However, the surprising findings recently reported in  \cite{SGFx} show that   error floors can be indeed   avoided by using hybrid SIC, where the SIC decoding order is opportunistically  chosen, as explained in the following.

Prior to  user scheduling, define a threshold for evaluating  the secondary users' channel conditions as follows:
\begin{align}\label{taux}
\tau  = \max\left\{0, \frac{|h_0|^2}{2^{R_0}-1}-\frac{1}{P}\right\}.
\end{align} 
 By using the threshold, the $M$ secondary users can be divided into two groups:
\begin{itemize}
\item Group 1, denoted by $\mathcal{S}_1$, contains the  users with strong channel conditions, i.e., $|h_n|^2>\tau$, and can support  the CSI-based SIC decoding order   only. If a user in $\mathcal{S}_1$ is scheduled, the base station will decode the secondary user's signal first, which yields the following   data rate:
\begin{align}
R^{1}_n =  \log\left(1+\frac{P|h_n|^2}{1+P|h_0|^2}\right),
\end{align}
for $n\in \mathcal{S}_1$.

\item Group 2, denoted by $\mathcal{S}_2$, contains the users which have relatively weak channel conditions, i.e., $|h_n|^2<\tau$, and can support either of the two SIC decoding orders.   If a user from  $\mathcal{S}_2$ is scheduled, its achievable data rate is $\log\left(1+\frac{P|h_n|^2}{1+P|h_0|^2}\right)$ for  CSI-based SIC, and $\log(1+|h_n|^2P)$ for QoS-based SIC. Therefore, the achievable data rate for a user in $\mathcal{S}_2$ is given by
{\small \begin{align}\nonumber
R^{2}_n =&  \max\left\{\log\left(1+\frac{P|h_n|^2}{1+P|h_0|^2}\right),\log(1+P|h_n|^2)\right\}\\  =&\log(1+P|h_n|^2) ,
\end{align}}
for $n\in \mathcal{S}_2$, where $\max\{a,b\}$ denotes the maximum of $a$ and $b$. 
\end{itemize}

With global CSI at the base station, the user with the maximum  achievable data rate  is admitted to the channel. Hence, the achievable data rate of the admitted user is given by
 \begin{align}
R^* = \max\left\{ \max\left\{R^{1}_n , \forall n\in\mathcal{S}_1\right\},  \max\left\{R^{2}_n , \forall n\in\mathcal{S}_2\right\}\right\}.
\end{align} 
 
 {\it Remark 3:} The considered  SIC scheme can be viewed as a   hybrid version of CSI- and QoS-based SIC, since both decoding orders can be used. Intuitively, one would expect that   this straightforward combination cannot avoid an outage probability   error floor, since both   SIC decoding orders suffer individually  from this drawback. However,  contrary to intuition, this simple hybrid scheme  can indeed eliminate  those error floors, and ensure that the outage probability  goes to zero for high SNR, as explained in the following subsection.

{\it Remark 4:} We note that NOMA with hybrid   SIC    can be implemented without   global CSI   at the base station \cite{SGFx}. In particular, distributed contention control can be applied, where the users in $\mathcal{S}_1$ choose  their backoff time inversely proportional to $R^1_n$, and the users in $\mathcal{S}_2$ choose their backoff time inversely proportional to $R^2_n$. As such,  the user with $R^*$ can be granted access  in a distributed manner.

\subsection{The Performance  of Hybrid SIC with Adaptive Decoding Order  }  
The outage probability experienced by the admitted user is given by
\begin{align}
{\rm P}^o =& {\rm P}\left(R^*<R_s\right)
\\\nonumber
= &{\rm P}\left(R^{1}_n<R_s , \forall n\in\mathcal{S}_1, R^{2}_m <R_s, \forall m\in\mathcal{S}_2\right).
\end{align} 
By using the assumption  that the users's channels are ordered as $|h_1|^2\leq \cdots\leq    |h_M|^2$,  the outage probability can be upper bounded as follows:
 \begin{align}\nonumber 
{\rm P}^o  
= &{\rm P}\left(R^{1}_n<R_s , \forall n\in\mathcal{S}_1, R^{2}_m <R_s, \forall m\in\mathcal{S}_2, |\mathcal{S}_2|>0\right)\\ \nonumber &+{\rm P}\left(R^{1}_n<R_s , \forall n\in\mathcal{S}_1, |\mathcal{S}_2|=0\right)\\ \label{eq18}
\leq &{\rm P}\left( \log(1+|h_m|^2P)<R_s, \forall m\in\mathcal{S}_2, |\mathcal{S}_2|>0\right)\\  \label{eq19}&+\underset{Q_0}{\underbrace{{\rm P}\left(R^{1}_M<R_s ,  |h_1|^2>\tau\right)}},
\end{align} 
where $|\mathcal{S}|$ denotes the size of set $\mathcal{S}$. It is straightforward to show that the probability in \eqref{eq18} goes to zero when $P\rightarrow \infty$, but the probability  in \eqref{eq19}, denoted by $Q_0$, is less straightforward to analyze.  

Since  $R^{CSI}_M=R^{1}_M$,   the outage probability for   CSI-based SIC  can be rewritten as  ${\rm P}^{CSI}={\rm P}\left(R^{1}_M<R_s  \right)$, which is quite similar to  $Q_0$ in \eqref{eq19}, where the only difference is that there is an extra term $|h_1|^2>\tau$ in $Q_0$.   At first glance,  the event   $|h_1|^2>\tau$ is trivial, and   an error floor should still exist for $Q_0$, similar to that for ${\rm P}^{CSI}$. However,  the additional term $|h_1|^2>\tau$ introduces  a hidden constraint which effectively eliminates   any error floors, as explained in the following.  
 $Q_0$ can be first rewritten  as follows:
 \begin{align}\nonumber
 Q_0 =& 
{\rm P}\left( |h_M|^2  <\frac{(1+P|h_0|^2)(2^{R_s}-1)}{P}  , |h_1|^2>\tau\right) .
\end{align} 
The fact that the upper bound on $|h_M|^2$ needs to be larger than the lower bound on $|h_1|^2$ results in the following   constraint:
\begin{align}
 \frac{(1+P|h_0|^2)(2^{R_s}-1)}{P}  >\tau .
\end{align}
This hidden constraint can be rephrased as follows:
\begin{align}
  |h_0|^2<\frac{2^{R_s}}{P\left(\frac{1}{2^{R_0}-1} - (2^{R_s}-1)\right)}   ,
\end{align}
where we assume    that $\tau>0$  and $\left(2^{R_s}-1\right)\left(2^{R_0}-1\right)<1$.
Therefore, $Q_0$ can be upper bounded as follows:
 \begin{align}\label{lowereb1}
 Q_0 \leq 
{\rm P}\left( |h_0|^2<\frac{2^{R_s}}{P\left(\frac{1}{2^{R_0}-1} - (2^{R_s}-1)\right)} \right) ,
\end{align} 
which   goes to zero when $P\rightarrow \infty$. Combining    \eqref{eq18}, \eqref{eq19}, and \eqref{lowereb1}, it is straightforward to show that an error floor for $P^o$ does not exist. We note that  hybrid SIC   can also  realize a   multi-user diversity gain of $M$, as shown  in detail in \cite{SGFx}.

\begin{figure}[t] \vspace{-1em}
\begin{center}\subfigure[ The Admitted User's Outage Probability ]{\label{fig2a}\includegraphics[width=0.4\textwidth]{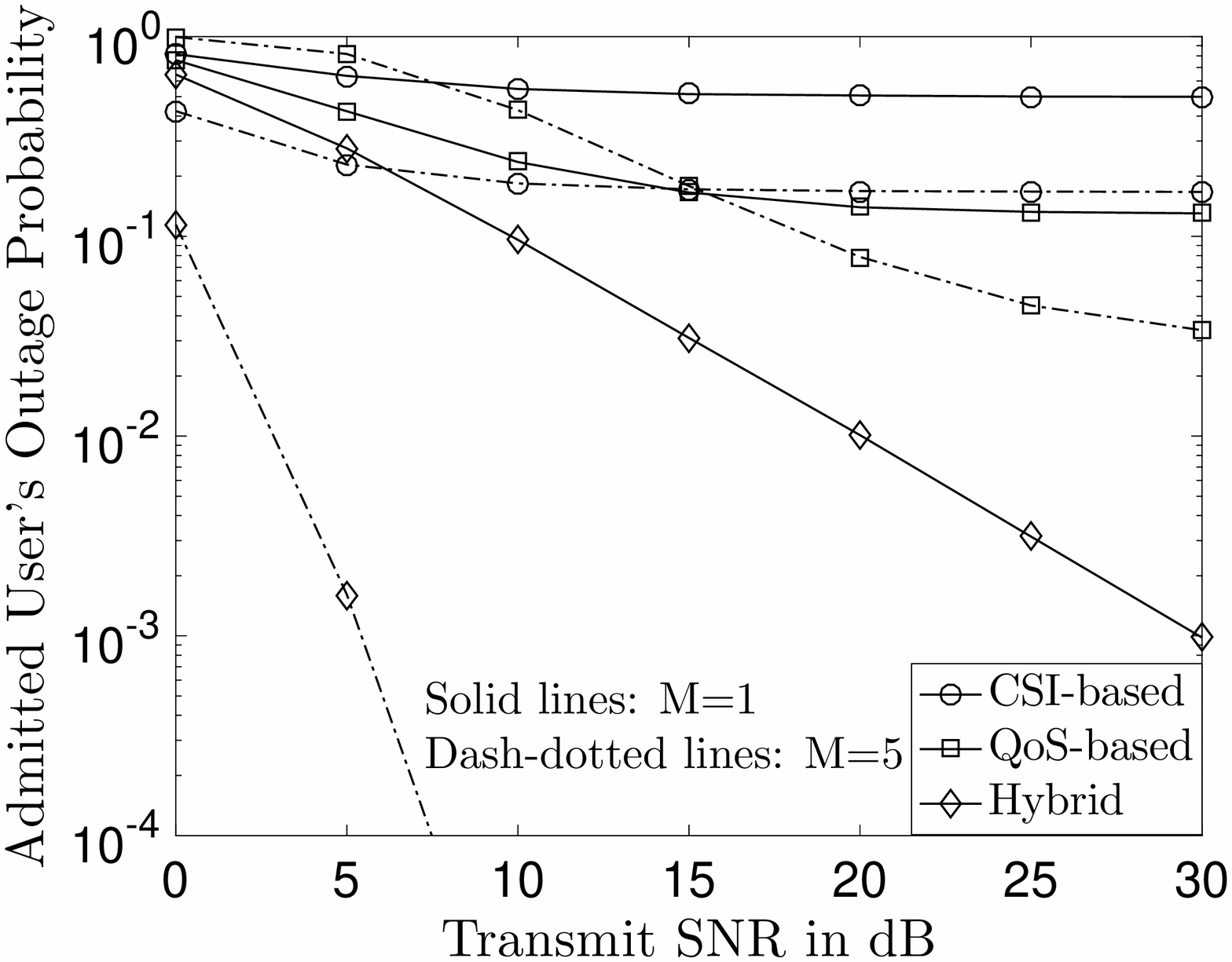}}
\subfigure[Sum Rate Gain]{\label{fig2b}\includegraphics[width=0.4\textwidth]{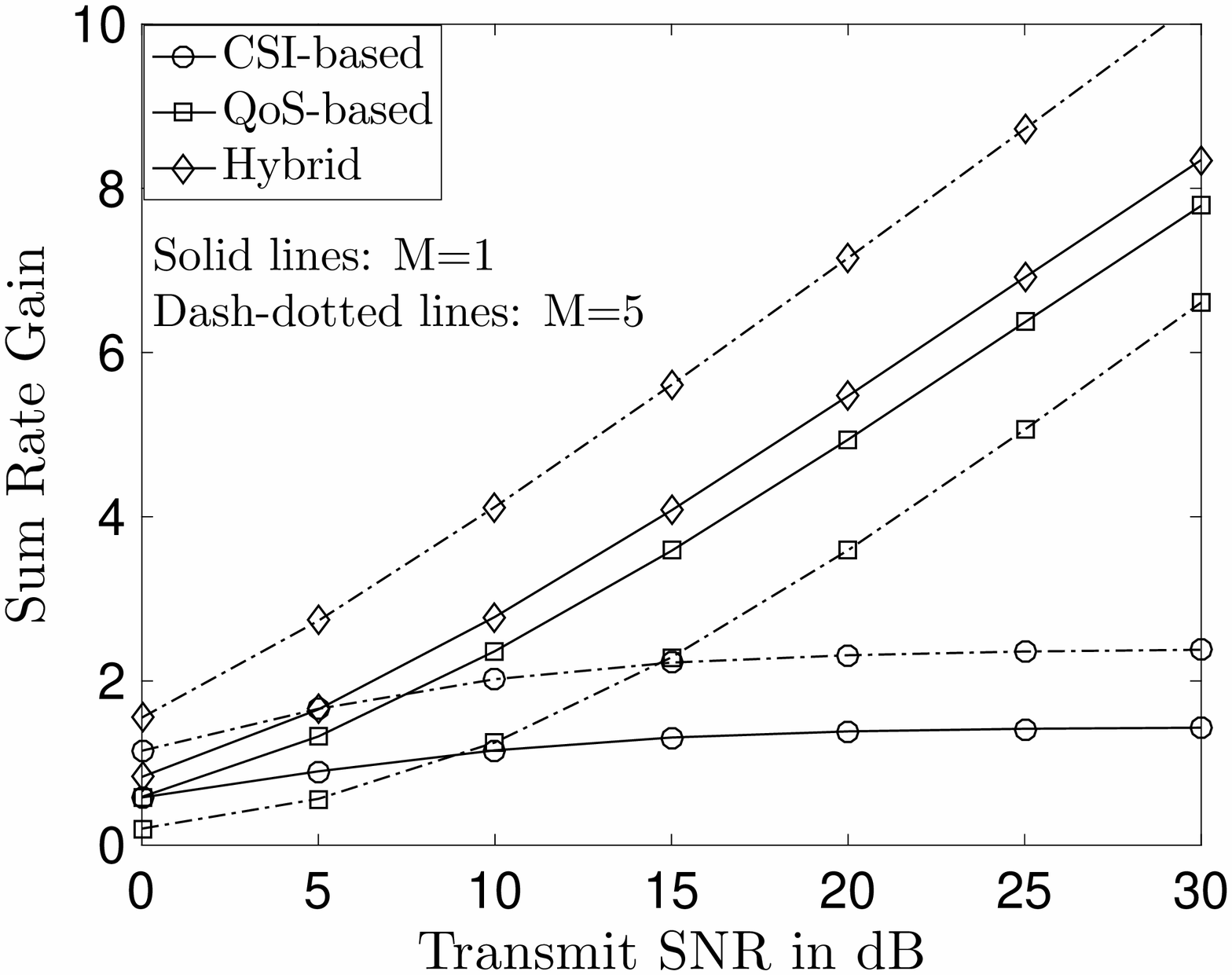}} \vspace{-1em}
\end{center} 
 \caption{The performance achieved by NOMA transmission with the three types of SIC. I.i.d. Rayleigh fading is assumed for the users' channel gains.   $R_0=0.2$ bits/s/Hz, and $R_s=1$ bits/s/Hz.    }\label{fig 2}\vspace{-1.5em}
\end{figure}
\subsubsection*{Numerical Studies} 
In Fig. \ref{fig 2}, the performance of NOMA with hybrid SIC  is demonstrated by using computer simulations, where       QoS-  and CSI-based SIC  are used as   benchmark schemes. In Fig. \ref{fig2a}, the outage performance achieved by the three NOMA schemes is shown. The figure demonstrates  that hybrid SIC always outperforms the CSI and QoS based SIC. More importantly,   hybrid SIC   can avoid outage probability error floors, as shown in Fig. \ref{fig2a}. Furthermore,  the fact that the slope of the curve for  NOMA with hybrid SIC   is increased by increasing $M$ indicates that  hybrid SIC   can effectively   exploit   multi-user diversity, i.e., inviting more users to participate in NOMA transmission improves transmission robustness. On   the contrary, the curves for   the  two benchmarking schemes exhibit error floors, and increasing $M$ degrades the performance of NOMA with QoS-based SIC, particularly at low SNR.    

All three NOMA schemes   ensure that ${\rm U}_0$ experiences the same  QoS as   in   OMA, which means that the admitted secondary user's data rate is the sum rate gain of NOMA over OMA.  Fig. \ref{fig2b} illustrates  the sum rate gains offered by the three NOMA schemes. In particular, the figure demonstrates that NOMA with hybrid SIC  always achieves  the largest performance gain  among the three schemes. In addition, increasing $M$   improves the performance gain offered by hybrid SIC. An interesting observation in Fig. \ref{fig2b} is that the performance of NOMA with QoS-based SIC   is degraded by increasing $M$, since $R^{QoS}_1$ shown in \eqref{qosrate} is a function of $|h_1|^2$ and the value of  $|h_1|^2$  is reduced as $M$ grows. 

\vspace{-0.5em}
 \section{Conclusions}
In the first part of this invited paper,  we have reviewed   the state of the art and   recent progress regarding the selection of the SIC decoding order for NOMA systems. In particular,    CSI-based SIC   was introduced first, and then   QoS-based SIC   was described.  The limitations of the two predefined  SIC decoding order selection schemes  were illustrated, and used as the motivation for  the recently proposed  hybrid SIC scheme with adaptive decoding order.   A comparison of these SIC schemes   was provided, and the reasons behind their performance differences  were also explained in detail.   

The recent findings in \cite{SGFx} are particularly exciting. Using the simple trick of switching between the  possible   SIC decoding  orders, hybrid SIC yields  a significant performance gain, i.e., removing the outage probability error floors, which cannot be achieved by CSI- and QoS-based SIC. These findings are particularly valuable given the fact that most existing works on NOMA adopt  a prefixed SIC decoding  order based on either the users' CSI or their  QoS requirements. Therefore, a natural question is whether these recent findings can be extended to other types of NOMA communication scenarios, which will be discussed in the second part of this invited paper. 

\vspace{-1em}

     \bibliographystyle{IEEEtran}
\bibliography{IEEEfull,trasfer}

   \end{document}